\documentstyle[12pt]{article}

\begin{document}
\bigskip

\title{Vacuum Condensates in the Global Color Symmetry Model}

\bigskip

\author{{Hong-shi Zong$^{1,2}$},  Xiao-fu L\"{u}$^{1,2,3}$, Jian-zhong Gu$^{1,2,4}$ \\
Chao-Hxi Chang$^{1,2}$ and En-guang Zhao$^{1,2}$ \\
\normalsize {$^{1}$ CCAST(World Laboratory), P.O. Box 8730,
Beijing 100080, P. R. China} \\
\normalsize{$^{2}$ Institute of Theoretical Physics, Academia Sinica, P.O. Box 2735,} \\
\normalsize{Beijing 100080, P. R. China}\thanks{Mailing Address}\\
\normalsize{$^{3}$  Department of Physics, Sichuan University,
Chengdu 610064, P. R. China}\\
\normalsize{$^{4}$  Max-Planck-Institut f\"{u}r Kernphysik, Postfach 103980,
D-69029 Heidelberg, Germany}} 

\date{}
\maketitle
\vspace*{0.5cm}

\begin{center}
{\large \bf Abstract}\\
\vspace{.5cm}
\begin{minipage}{13cm}

\parindent=18pt

{\small Based on the quark propagator in the instanton dilute liquid
approximation, we calculate analytically the quark condensate $<\bar{q}q>$, 
the mixed quark gluon condensate
$g_{s}<\bar{q}G_{\mu\nu}\sigma^{\mu\nu}q>$ and the four
quark condensate $<\bar{q} \Gamma q\bar{q} \Gamma q>$ at the mean field level
in the framework of global color symmetry model. The numerical calculation
shows that the values of these condensates are compatible with the ranges
determined by other
nonperturbative approaches. Moreover, 
we find that for nonlocal four quark condensate
the previous
vacuum saturation assumption is not a good approximation 
even at the mean field level.} 
\end{minipage}
\end{center}

\vspace*{1.5cm}

PACS Numbers. 24.85.+p, 12.38.Lg; 12.38; 11.15Pg

\newpage

\parindent=20pt

\parskip=5mm

The non-perturbative structure of QCD vacuum is characterized by the 
vacuum matrix elements of various singlet combination of quark and gluon
fields, especially by such condensates as the quark condensate $<\bar{q}q>$, the mixed
quark gluon condensate $g_{s}<\bar{q}G_{\mu\nu}\sigma^{\mu\nu}q>$ and the four
quark condensate $<\bar{q} \Gamma q\bar{q} \Gamma q>$. These condensates
would vanish in a perturbative vacuum but would not in the
non-perturbative QCD vacuum, and are of essence for describing the physics
of strong interaction at low and intermediate energies. Naturally, it is
interesting to determine the vacuum values at least of the lower dimensional
quark and gluonic operators.

Previous studies of the condensates
of $<\bar{q}q>$, $g_{s}<\bar{q}G_{\mu\nu}\sigma^{\mu\nu}q>$ and
$<\bar{q} \Gamma q\bar{q} \Gamma q>$
include QCD sum rules  where the condensates were treated as fit parameters in the
analysis of various hadron channels [1-3], quenched lattice QCD[4], the
instanton liquid model[5] and the effective quark-quark interaction model[6].
In the present letter we shall investigate the vacuum condensates in the
framework of the global color symmetry model(GCM)[7-9] which is based on 
effective quark-quark interaction and can be defined through a truncation
of QCD as follows. The generating functional for QCD in the Euclidean
metric is 
\begin{equation}
Z[\bar{\eta}, \eta] = \int {\cal{D}}\bar{q} {\cal{D}} q {\cal{D}}
\bar{\omega} {\cal{D}} \omega {\cal{D}} A \exp\left\{-S -S_{gf} -S_{g}
+ \int d^4 x (\bar{\eta} q + \bar{q} \eta)\right\} \, ,
\end{equation}
where
$$ S = \int d ^4 x \left\{ \bar{q} \left[ \gamma _{\mu} \left(\partial_{\mu}
- i g \frac{\lambda ^{a}}{2} A^{a}_{\mu} \right) \right] q + \frac{1}{4}
F_{\mu \nu}^{a} F_{\mu \nu}^{a} \right\}, $$
and $S_{gf}$, $S_g$ are the gauge-fixing and ghost actions.
As a bilocal field $B^{\theta}(x,y)$ being introduced, the generating
functional can be given as
\begin{eqnarray}
Z[\bar{\eta}, \eta] & = & \exp {\left[W_1 \left(ig \frac{\delta}{\delta
\eta(x)} \frac{\lambda^{a}}{2} \gamma_{\mu} \frac{\delta}{\delta
\bar{\eta}(x)} \right) \right] }  \nonumber \\  & & \times 
\int {\cal{D}}\bar{q} {\cal{D}} q {\cal{D}} B^{\theta}(x,y)
\exp\left\{-S [ \bar{q}, q, B^{\theta}(x,y)] + \int d^4 x ( \bar{\eta} q +
\bar{q} \eta)\right\} \, , 
\end{eqnarray}
where
$$ W_1 [ J_{\mu}^{a} ] = \sum_{n=3}^{\infty} \frac{1} {n !} \int d x_1
\cdots d x_n D_{\mu _1 \cdots \mu_n}^{a_1 \cdots a_n} (x_1,\cdots, x_n)
\Pi_{i=1}^{n} J_{\mu _i}^{a_i} (x_i) \, , $$
and 
$$ S[\bar{q}, q, B^{\theta}(x,y)] = \int \int d^4 x d^4 y \left[ \bar{q}(x)
G^{-1}(x,y;[B^{\theta}]) q(y) + \frac{B^{\theta}(x,y) B^{\theta}(y,x)}
{2 g^2 D(x-y)} \right] \; , $$
with
\begin{equation}
G^{-1}(x,y; [B^{\theta}]) = \gamma \cdot \partial \delta(x-y) +
\frac{1}{2} \Lambda^{\theta} B^{\theta}(x,y) \, .
\end{equation}
Here $\Lambda^{\theta} = K^{a} C^{b} F^{c}$ is determined by Fierz
transformation in the color, flavor and Lorentz space [7], and $g^2D(x-y)$ is the
effective gluon propagator in GCM.

Neglecting $W_1[J_{\mu}^{a}]$ we obtain the GCM generating functional 
\begin{equation}
Z_{GCM}[\bar{\eta}, \eta] = \int {\cal{D}}\bar{q} {\cal{D}} q {\cal{D}}
B^{\theta}(x,y) \exp\left\{-S [ \bar{q}, q, B^{\theta}(x,y)] +
\int d^4 x ( \bar{\eta} q + \bar{q} \eta)\right\} \, . 
\end{equation}

Performing the functional integral over ${\cal{D}} \bar{q}$ and
${\cal{D}} q$ in Eq. (4), we have
\begin{equation}
Z_{GCM}[\bar{\eta}, \eta] = \int {\cal{D}} B^{\theta}(x,y) \exp(-S [ \bar{\eta},
\eta, B^{\theta}(x,y)] ) \, , 
\end{equation}
where
\begin{eqnarray}
S_[\bar{\eta}, \eta, B^{\theta}(x,y)] & = & - \mbox{Tr} \ln \left[\rlap/
\partial \delta (x-y) + \frac{1}{2} \Lambda^{\theta} B^{\theta}(x,y) \right]
\nonumber \\
& & + \int\int \left[\frac{B^{\theta}(x,y)
B^{\theta}(y,x)}{2 g^2 D(x-y)}
+ \bar{\eta}(x) G(x,y;[B^{\theta}])\eta(y) \right] .
\end{eqnarray}
The saddle-point of the action is defined as
$\left.\delta S[\bar{\eta}, \eta, B^{\theta}(x,y)]/\delta B^{\theta}(x,y)
\right\vert _{\eta =\bar{\eta}
= 0} =0$ and is given by
\begin{equation}
B^{\theta}_{0}(x-y)=g^2 D(x-y) tr_{\gamma C}[\Lambda^{\theta} G_{0}(x-y)],
\end{equation}
where $G_{0}$ stands for $G[B^{\theta}_{0}]$. These configurations are
related to vacuum condenstates and
provide self-energy dressing of the quarks through the definition
$\Sigma(p)\equiv \frac{1}{2} \Lambda^{\theta}B^{\theta}_{0}(p)=i\gamma\cdot
p[A(p^2)-1]+B(p^2)$. The self energy functionals $A(p^2)$ and $B(p^2)$ are
determined by the rainbow Dyson-schwinger equations[10]
\begin{equation}
[A(p^2)-1]p^2=\frac{8}{3}\int \frac{d^{4}q}{(2\pi)^4}g^2 D(p-q)
\frac{A(q^2)p\cdot q}{q^2A^2(q^2)+B^2(q^2)},
\end{equation}
\begin{equation}
B(p^2)=\frac{16}{3}\int \frac{d^{4}q}{(2\pi)^4}g^2 D(p-q)
\frac{B(q^2)}{q^2A^2(q^2)+B^2(q^2)}.
\end{equation}
This dressing comprises the notion of constituent quarks by providing 
a mass $M(p^2)=B(p^2)/A(p^2)$, reflecting a vacuum configuration with
dynamically broken chiral symmetry. Because the form of the gluon propagator
$g^2D(s)$ in the IR region remains missing. One often uses model forms as input
in the previous studies of the rainbow Dyson-Schwinger equations [6-10]. On the
other hand, much evidence of strong
interaction phenomena is recently in favour of the instanton structure
of the vacuum in QCD [11-14]. In particlar,
the complex configuration of the vacuum of QCD with different topological
winding numbers can be constructed in the model of an instanton dilute
liquid [11,12]. Lattice QCD [15] calculations are also consistent with the above
point of view. Therefore, contrary to the previous philosophy, we use the
quark propagator in the model of the instanton dilute liduid instead of the
gluon propagator as input (more detials can be seen in Ref.[16]). In Ref.[16],
we showed that the self energy function $A(q^2)=1$ and $B(q^2)$
is equivalent to the dynamical mass $m(q^2)$ of quarks in the
instanton dilute liquid approximation, i.e.
\begin{equation}
B(q^2)=m(q^2) = \frac{\epsilon \bar{\rho} }{6} q^2 \varphi^{2}(q) \; ,
\end{equation}
where $\varphi(q) = \pi {\bar{R}}^2 \frac{d}{dz}[I_0(z) K_0(z)
-I_1(z) K_1(z)]$ with $z= \frac{\vert q \vert \bar{R} }{2}$. 
$I_n(z) \; (K_n(z))$ \linebreak $(n=0,1)$ are the first (second) kind 
modified Bessel functions of order $n$. $\bar{\rho} \approx $
(200~MeV)$^4$ is the average density of the instantons, $\bar{R}= 
\frac{1}{3}~{\mbox{fm}}$ the average radius of the instantons, and
$\epsilon$ a constant (85~MeV)$^{-1}$. 

Since the self energy function $A(q^2)$ and $B(q^2)$ have been determined in the
instanton dilute liquid approximation 
we can calculate the vacuum
condensates by the above saddle-point expansion, that is, we shall work at
the mean field level.

According to the definition of the GCM generating functional it is now straightforward
to calculate the vacuum expectation value of any quark operator of the form
\begin{equation}
O_{n}\equiv(\bar{q}_{j1}\Lambda^{(1)}_{j1i1}q_{i1})
(\bar{q}_{j2}\Lambda^{(2)}_{j2i2}q_{i2})\cdots
(\bar{q}_{jn}\Lambda^{(n)}_{jnin}q_{in})
\end{equation}
in the mean field vacuum. Here the $\Lambda^{(i)}$ stands for an operator in
the Dirac, flavor and color space.

Taking the appropriate value of derivatives of Eq.(5) with respect to external sources
$\eta_{i}$ and $\bar{\eta_{j}}$ 
(putting $\eta_{i}=\bar{\eta}_{j}=0$ [17]),
we have
\begin{equation}
<:O_{n}:>=(-)^{n}\sum_{p}[(-)^{p}\Lambda^{(1)}_{j1i1}
\cdots\Lambda^{(n)}_{jnin}(G_0)_{i1jp(1)}\cdots(G_0)_{injp(n)}]
\end{equation}
where p stands for a permutation of the n indices. In particular, we obtain
the nonlocal quark condensate $<:\bar{q}(x)q(0):>$
\begin{eqnarray}
<:\bar{q}(x)q(0):>_{\mu}&=&(-)tr_{\gamma C}[G_{0}(x,0)]\nonumber\\
&=&(-4N_{c})\int^{\mu}_{0}\frac{d^4 p}{(2\pi)^4}\frac{B(p^2)}
{p^2A^2(p^2)+B^2(p^2)}e^{ipx}\nonumber\\
&=&(-)\frac{12}{16\pi^2}\int^{\mu}_{0}ds s\frac{B(s)}
{sA^2(s)+B^2(s)}[2\frac{J_1(\sqrt{sx^2})}{\sqrt{sx^2}}],
\end{eqnarray}
where $\mu$ is the renormalization scale which we choose to be 1 $GeV^2$.
At x=0 the expression for the local condensate $<:\bar{q}q:>$ is reproduced
\begin{equation}
<:\bar{q}q:>_{\mu}=(-)tr_{\gamma C}[G_{0}(x,0)]|_{x=0}=(-)\frac{12}{16\pi^2}\int^{\mu}_{0}ds s
\frac{B(s)}{sA^2(s)+B^2(s)}.
\end{equation}
The nonlocality $g(x^2)$ can be obtained immediately by dividing Eqs.(13)
by Eq.(14).

\begin{tabular}{lllllllllllr}
\multicolumn{12}{c}{Table. I. The nonlocal quark condensate
$g(x)=<:\bar{q}(x)q(0):>/<:\bar{q}(0)q(0):>$.}\\ \hline\hline
$x^2(GeV^2)$ &0.0 &2.0 &4.0 &6.0 &8.0 &10  &12  &14  &16  &18  &20 \\ \hline
$g(x^2)$     &1.0 &0.89&0.80&0.71&0.64&0.57&0.50&0.45&0.40&0.35&0.31\\ \hline\hline       
\end{tabular}

In Table I we display the nonlocal quark condensate g(x) versus $x^{2}$. By 
comparing it with the corresponding results mentioned in the Ref.[18] we
demonstrate that the nonlocal quark condensate is very robust 
in our approach or the method of model gluon propagators.

Another important vacuum condensate following from Eq.(11) is the nonlocal four quark
condensate in the mean filed vacuum. In the case of $\Lambda^{(1)}$
=$\Lambda^{(2)}$=$\gamma_{\mu}\frac{\lambda^a_{C}}{2}$ we find 
\begin{eqnarray}
& &<:\bar{q}(x)\gamma_{\mu}\frac{\lambda^a_{C}}{2}q(x)\bar{q}(0)
\gamma_{\mu}\frac{\lambda^a_{C}}{2}q(0):>_{\mu}\nonumber\\
&=&-tr_{\gamma C}[G_{0}(0,x)\gamma_{\mu}\frac{\lambda^a_{C}}{2}
G_{0}(x,0)\gamma_{\mu}\frac{\lambda^a_{C}}{2}]
+tr_{\gamma C}[G_0(x,x)\gamma_{\mu}\frac{\lambda^a_{C}}{2}]
tr_{\gamma C}[G_{0}(0,0)\gamma_{\mu}\frac{\lambda^a_{C}}{2}]\nonumber\\
&=&(-)\int^{\mu}_{0}\int^{\mu}_{0}
\frac{d^4p}{(2\pi)^4}\frac{d^4q}{(2\pi)^4}e^{ix\cdot(p-q)}
\left[4^3\frac{B(p^2)}{A^2(p^2)p^2+B^2(p^2)}\frac{B(q^2)}
{A^2(q^2)q^2+B^2(q^2)}\right.\nonumber\\
& &\left.+ 2\times 4^2\frac{A(p^2)}{A^2(p^2)p^2+B^2(p^2)}\frac{A(q^2)}
{A^2(q^2)q^2+B^2(q^2)}p\cdot q\right].
\end{eqnarray}
Similarly, at x=0 the expression for the local four quark condensate
$<:\bar{q}\gamma_{\mu}\frac{\lambda^a_{C}}{2}q\bar{q}
\gamma_{\mu}\frac{\lambda^a_{C}}{2}q:>$ is recovered:
\begin{equation}
<:\bar{q}\gamma_{\mu}\frac{\lambda^a_{C}}{2}q\bar{q}
\gamma_{\mu}\frac{\lambda^a_{C}}{2}q:>_{\mu}=(-4^3)[\int^{\mu}_{0}\frac{d^4p}{(2\pi)^4}
\frac{B(p^2)}{A^2(p^2)p^2+B^2(p^2)}]^2=(-)\frac{4}{9}<:\bar{q}q:>^2,
\end{equation}
namely, for the local four quark condensate, our result is consistent
with the vacuum saturation assumption of Ref.[1]. However, if we consider the
nonlocal four quark condensate, it should be noted that the contribution of
the second term of right-hand of Eq.(15) can not be neglected.

As to the mixed condensate $g_{s}<\bar{q}G_{\mu\nu}\sigma^{\mu\nu}q>$, we
can use the method described by Ref.[6] to obtain the mixed condensate in
Minkowski space
\begin{equation}
g_{s}<\bar{q}G_{\mu\nu}\sigma^{\mu\nu}q>_{\mu}
=(-)(\frac{N_{c}}{16\pi^2})\{\frac{27}{4}
\int_{0}^{\mu}dss\frac{B[2A(A-1)s+B^2]}
{A^2s+B^2}+12\int_{0}^{\mu}dss^2\frac{B(2-A)}{A^2s+B^2}\}.
\end{equation}

In Table II we display the result for $<\bar{q}q>$ and
$g_{s}<\bar{q}G_{\mu\nu}\sigma^{\mu\nu}q>$ in our approach and compare it
with the corresponding values which were obtained from other nonperturbative approaches 
(QCD sum rules[3], quenched lattice QCD[4], the instanton liquid model[5]
and the effective quark-quark interaction model[6]).

\begin{tabular}{llr} 
\multicolumn{3}{c}{Table. II, $<\bar{q}q>$ and $g_{s}<\bar{q}G_{\mu\nu}\sigma^{\mu\nu}q>$
in different non-perturbative approaches}\\ \hline\hline
                 & (--)$<\bar{q}q>^{\frac{1}{3}}$ & (--)$<g_{s}\bar{q}\sigma Gq>^{\frac{1}{5}}$ \\ \hline
present work        & 207 MeV                     & 415 MeV              \\
QCD sum rules[3] & 210 --230 MeV                 & 375 --395 MeV  \\
quenched lattice[4]& 225 MeV                   & 402 --429 MeV   \\
instantion liquid model[5] & 272 MeV              & 490 MeV \\
effective quark-quark interaction model[6] & 150 --180 MeV & 400 --460 MeV \\ \hline\hline
\end{tabular}

Table II shows that our results for $<\bar{q}q>$ and
$g_{s}<\bar{q}G_{\mu\nu}\sigma^{\mu\nu}q>$ are compatible with the ranges
obtained from other nonperturbative methods, especially from QCD sum
rules[3] and quenched lattice[4]. It should be noted that there is not any
free parameter in our model. In this model, the calculated 
masses, decay constants of the mesons $\pi$ and $\sigma$ and
decay width of $\sigma \rightarrow \pi\pi$ agree with experimental data
quite well[16].

In summary, based on the instanton dilute liquid approximation, we have
determined the quark condensate $<\bar{q}q>$, the mixed
quark gluon condensate $g_{s}<\bar{q}G_{\mu\nu}\sigma^{\mu\nu}q>$ and the four
quark condensate $<\bar{q} \Gamma q\bar{q} \Gamma q>$ at the mean field level
in the framework of GCM. The numerical calculation
have shown that our results are compatible with the ranges obtained from other
nonperturbative approaches. Futhermore, we have found that the contribution
of the second term of right-hand of Eq.(15) must be taken into account when one
calculates nonlocal four quark condensate. This implies that even at the mean
field level the previous vacuum
saturation assumption is not a good approximation for the 
nonlocal four quark condensate.

\noindent{\large \bf Acknowledgement}

This work was supported in part by the National Natural Science Foundation
of China.

\vspace*{2.cm}

\noindent{\large \bf References}
\begin{description}
\item{[1]} M. Shifman, A. Vainshtein and V. Zakharov, Nucl. Phys 
{\bf B147}, 385 (1979).
\item{[2]}  L. Reinders, H. Rubinstein and S. Yazaki, Phys. Rep.
{\bf 127}, 1 (1985).
\item{[3]} S. Narison, QCD Spectral Sum Rules (World Scientific, Singapore
, 1989), and references therein.
\item{[4]} M. Kremer and G. Schierholz,  Phys. Lett. {\bf B194}, 283 (1987).
\item{[5]} M. V. Polyakov and C. Weiss, Phys. Lett. {\bf B387}, 841 (1996). 
\item{[6]} T. Meissner, Phys. Lett. {\bf B405}, 8 (1997).
\item{[7]} R.~T. Cahill and C.~D. Roberts, Phys. Rev. {\bf D32}, 2419 (1985).
\item{[8]} P. C. Tandy, Prog. Part. Nucl. Phys. 39, 117 (1997).
\item{[9]} R.~T. Cahill, Nucl. Phys. {\bf A543}, 63c (1992).
\item{[10]} C. D. Roberts and A. G. Williams, Prog. Part. Nucl. Phys.{\bf 33},
477 (1994), and references therein.
\item{[11]} E.~V. Shuryak, Phys. Rep. {\bf 115}, 151 (1984); T. Schafer and
E. V. Shuryak. Rev. Mod. Phys. {\bf 70}, 323 (1998).
\item{[12]} D.~I. Dyakonov and X. Yu Petrov, Nucl. Phys. {\bf B272}, 457 (1986). 
\item{[13]} S. Takeuchi and M. Oka, Nucl. Phys. {\bf A547}, 283c (1992).
\item{[14]} H. Forkel and M. K. Banerjee, Phys. Rev. Lett.{\bf 71},484 (1993).
\item{[15]} M. C. Chu, J. M. Grandy, S. Huang and J. Negele, Phys. Rev.
{\bf D49}, 6039 (1994); J. Negele, ``Lattice QCD'' presented at the
International Summer School and Workshop on Nuclear QCD, Beijing,1995.
\vspace*{-3mm}
\item{[16]} Xiao-fu L\"{u}, Yu-xin Liu, Hong-shi Zong and En-guang Zhao,
Phys. Rev. {\bf C58}, 1195 (1998).
\item{[17]} J. Negele and H. Orland, Quantum Many-particles Systems
(Addison-Wesley 1988).
\item{[18]} L. S. Kisslinger and T. Meissner, Phys. Rev. {\bf C57},1528 (1998).
\end{description}

\end{document}